\documentclass[prl,twocolumn,superscriptaddress,nobibnotes,letterpaper]{revtex4-1}

\usepackage[pdftex]{graphicx}
\usepackage[pdftex]{epsfig}
\usepackage{amsmath}
\usepackage{amssymb}
\usepackage{amsfonts}
\usepackage{color}
\usepackage{wrapfig}
\usepackage{eucal}
\usepackage{hhline}
\usepackage{threeparttable}
\usepackage{supertabular}
\usepackage{multirow}
\usepackage{tabularx}

\usepackage{soul}

\usepackage{array} 
\usepackage[colorlinks=true, allcolors=blue]{hyperref} 

\newcommand{\opd}[2]{\mbox{$\hat{#1}_\text{#2}^{\dagger}$}}
\newcommand{\op}[2]{\mbox{$\hat{#1}_\text{#2}$}}

\newcommand{\ket}[2]{\mbox{$\rvert{#1}\rangle_\text{#2}$}}

\definecolor{cgreen}{rgb}{.1,.6,.1}
\definecolor{co}{rgb}{.1,.6,.6}
\definecolor{orange}{rgb}{.9,.4,.0}

\newcolumntype{C}[1]{>{\centering\arraybackslash}p{#1}}

\begin{document}
\pagenumbering{arabic}

\title{An optomechanical Bell test}\thanks{This work was published in Phys.\ Rev.\ Lett.\ \textbf{121}, 220404 (2018).}

\author{Igor Marinkovi\'{c}}\thanks{These authors contributed equally to this work.}
\affiliation{Department of Quantum Nanoscience, Kavli Institute of Nanoscience, Delft University of Technology, 2628CJ Delft, The Netherlands}
\author{Andreas Wallucks}\thanks{These authors contributed equally to this work.}
\affiliation{Department of Quantum Nanoscience, Kavli Institute of Nanoscience, Delft University of Technology, 2628CJ Delft, The Netherlands}
\author{Ralf Riedinger}
\affiliation{Vienna Center for Quantum Science and Technology (VCQ), Faculty of Physics, University of Vienna, A-1090 Vienna, Austria}
\author{Sungkun Hong}
\affiliation{Vienna Center for Quantum Science and Technology (VCQ), Faculty of Physics, University of Vienna, A-1090 Vienna, Austria}
\author{Markus Aspelmeyer}
\affiliation{Vienna Center for Quantum Science and Technology (VCQ), Faculty of Physics, University of Vienna, A-1090 Vienna, Austria}
\author{Simon Gr\"oblacher}
\email{s.groeblacher@tudelft.nl}
\affiliation{Department of Quantum Nanoscience, Kavli Institute of Nanoscience, Delft University of Technology, 2628CJ Delft, The Netherlands}


\begin{abstract}
Over the past few decades, experimental tests of Bell-type inequalities have been at the forefront of understanding quantum mechanics and its implications. These strong bounds on specific measurements on a physical system originate from some of the most fundamental concepts of classical physics -- in particular that properties of an object are well defined independent of measurements (realism) and only affected by local interactions (locality). The violation of these bounds unambiguously shows that the measured system does not behave classically, void of any assumption on the validity of quantum theory. It has also found applications in quantum technologies for certifying the suitability of devices for generating quantum randomness, distributing secret keys and for quantum computing. Here we report on the violation of a Bell inequality involving a massive, macroscopic mechanical system. We create light-matter entanglement between the vibrational motion of two silicon optomechanical oscillators, each comprising approx.\ $10^{10}$ atoms, and two optical modes. This state allows us to violate a Bell inequality by more than 4 standard deviations, directly confirming the non-classical behavior of our optomechanical system under the fair sampling assumption.
\end{abstract}

\maketitle

Bell's theorem~\cite{Bell1964} predicts that any local realistic theory is at variance with quantum mechanics. It was originally conceived to settle an argument between Einstein~\cite{EPR1935} and Bohr~\cite{Bohr1935} on locality in physics, and to investigate the axioms of quantum physics. First tests of the Clauser-Horne-Shimony-Holt (CHSH) inequality~\cite{Clauser1969}, an experimentally testable version of Bell's original inequality, were performed with photons from cascaded decays of atoms~\cite{Freedman1972,Aspect1981} and parametric down-conversion~\cite{Shih1988,Rarity1990,Kwiat1995}. Subsequent experiments reduced the set of assumptions required for the falsification of classical theories, closing, e.g., the locality~\cite{Weihs1998} and detection loopholes~\cite{Rowe2001}, first individually and recently simultaneously~\cite{Hensen2015,Giustina2015,Shalm2015,Rosenfeld2017}. In addition to the fundamental importance of these experiments, the violation of a Bell-type inequality has very practical implications -- in particular, it has become the most important benchmark for thrust-worthily verifying entanglement in various systems~\cite{Barreiro2013,Schmied2016}, including mesoscopic superconducting circuits~\cite{Ansmann2009}, for certifying randomness~\cite{Pironio2010,Bierhorst2018}, secret keys~\cite{Acin2007}, and quantum computing~\cite{Sekatski2018}.

While the standard form of quantum theory does not impose any limits on the mass or size of a quantum system~\cite{Schroedinger1935}, the potential persistence of quantum effects on a macroscopic scale seems to contradict the human experience of classical physics. Over the past years, quantum optomechanics has emerged as a new research field, coupling mechanical oscillators to optical fields. While these systems are very promising for quantum information applications due to their complete engineerability, they also hold great potential to test quantum physics on a new mass scale. Recent experiments have demonstrated quantum control of such mechanical systems, including mechanical squeezing~\cite{Wollman2015}, single-phonon manipulation~\cite{OConnell2010,Chu2017,Hong2017,Reed2017}, as well as entanglement between light and mechanics~\cite{Palomaki2013} and entanglement between two mechanical modes~\cite{Lee2011,Riedinger2018,Ockeloen-Korppi2018}. However, explaining the observed results in these experiments required assuming the validity of quantum theory at some level. A Bell test, in contrast, is a genuine test of non-classicality without quantum assumptions.

Here we report on the first Bell test using correlations between light and microfabricated mechanical resonators, which constitute massive macroscopic objects, hence verifying non-classical behavior of our system without relying on the quantum formalism. Bell-tests do not require assumptions about the physical implementation of a quantum system such as the dimension of the underlying Hilbert space or the fundamental interactions involved in state preparation and measurement~\cite{vanEnk2007}. The violation of a Bell-inequality is hence the most unambiguous demonstration of entanglement with numerous important implications. From a fundamental perspective, the robust entanglement between flying optical photons and a stored mechanical state rules out local hidden-variables, which can be used for further tests of quantum mechanics at even larger mass scales~\cite{CapraraVivoli2016,Hofer2016}. From an application perspective, the presented measurements also imply that optomechanics is a promising technique to be used for quantum information processing tasks including teleportation, quantum memories and the possibility of quantum communication with device-independent security~\cite{Acin2007}.

\begin{figure}[t!]
	\includegraphics[width=1.\columnwidth]{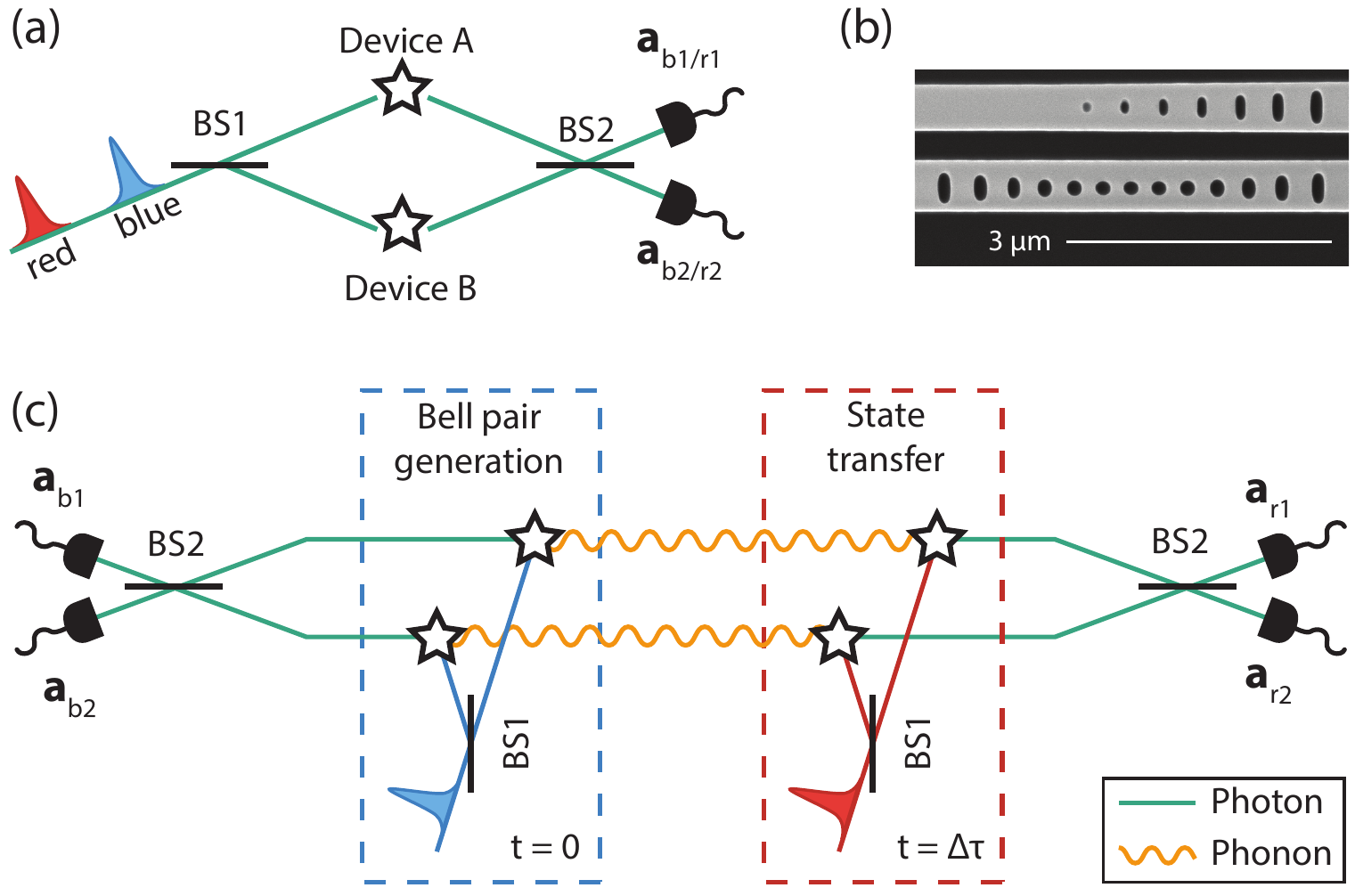}
	\caption{(a) Schematic of the setup:\ blue detuned drive pulses interact with the mechanical resonators (devices A \& B) producing entangled photon-phonon pairs. The light-matter entanglement is in the path basis (A or B), corresponding to the device in which the Stokes scattering event took place. The generated photons are detected in single-photon detectors giving the measurement results $a_{\mathrm{b1}}$ and $a_{\mathrm{b2}}$. The detection of the phonons is done by transferring their states to another optical mode by using a red drive after some time $\Delta\tau$ and subsequently obtaining the results $a_{\mathrm{r1}}$ and $a_{\mathrm{r2}}$. Note that for technical reasons the photons created by the blue and red drives are detected on the same pair of detectors, but with a time delay $\Delta\tau=200$~ns. Therefore we have time-separation of the two parties of the Bell test instead of space-separation (as commonly done). BS1/2 represent beamsplitter 1/2. (b) Scanning electron microscope image of one of the optomechanical devices, represented with a star-symbol in (a) and (c), next to the coupling waveguide (top). (c) Illustration of our experimental sequence: one party of the Bell test measures in which detector path the Stokes photon is found at time $t=0$, while the other performs the same measurement for the anti-Stokes photon after a time $t=\Delta\tau$. We probe their correlations in order to violate the CHSH inequality. Since the two photons never interacted directly (only through the mechanics), the observed correlations are a direct consequence of the correlations between the Stokes photons and phonons.}\label{fig:1}
\end{figure}

The optomechanical structures used in this work are two photonic crystal nanobeams on two separate chips. They are designed to have an optical resonance in the telecom band that is coupled to a co-localized, high-frequency mechanical mode~\cite{Chan2012}. Each device is placed in one of the arms of an actively stabilized fiber interferometer (see \cite{Riedinger2018} and SI for additional details). The resonators are cryogenically cooled close to their motional ground state inside a dilution refrigerator. Our entanglement creation and verification protocol consists of two optical control pulses that give rise to linearized optomechanical interactions, addressing the Stokes and anti-Stokes transitions of the system (see Figure~\ref{fig:1}). Both types of interactions result in scattered photons that are resonant with the cavity and can be efficiently filtered from the drive beams before being detected by superconducting nanowire single photon detectors (SNSPDs).

A blue detuned, $\sim$40~ns long laser pulse with frequency $\nu_\mathrm{b}=\nu_\mathrm{o}+\nu_\mathrm{m}$ ($\nu_\mathrm{o}$ optical resonance, $\nu_\mathrm{m}$ mechanical resonance) generates photon-phonon pairs. The interaction in this case is described by $\op{H}{b} = - \hbar g_0 \sqrt{n_\mathrm{b}} \opd{a}{}\opd{b}{}+\textrm{h.c.}$, with the intracavity photon number $n_\mathrm{b}$, the optomechanical single photon coupling $g_0$ and the optical (mechanical) creation operators $\opd{a}{}$ ($\opd{b}{}$). This correlates the number of mechanical and optical excitations in each of the arms of the interferometer as
\begin{eqnarray}
\ket{\psi}{}\propto (\ket{00}{om}+\epsilon\ket{11}{om}+\mathcal{O}(\epsilon^2)),
\end{eqnarray}
where $\mathrm{o}$ denotes the optical and $\mathrm{m}$ the mechanical mode, while $p=\epsilon^2$ is the excitation probability. For small $p\ll1$, states with multiple excitations are unlikely to occur, and can therefore be neglected in the statistical analysis. Driving the devices simultaneously and post-selecting on trials with a successful detection of both the Stokes-photon and the phonon, we approximate the combined state as
\begin{eqnarray}
\ket{\Psi}{}&=&\frac{1}{\sqrt{2}}(\ket{11}{A}\ket{00}{B}+e^{i\phi_\mathrm{b}}\ket{00}{A}\ket{11}{B})\nonumber\\
&=&\frac{1}{\sqrt{2}}(\ket{AA}{om}+e^{i\phi_\mathrm{b}}\ket{BB}{om}),
\end{eqnarray}
again neglecting higher order excitations. Here $\phi_\mathrm{b}$ is the phase difference that the blue drives acquire in the two interferometer paths A and B, including the phase shift of the first beam splitter. Expressing the state in a path basis $\ket{A}{x}=\ket{10}{AB}$, where $\mathrm{x}$ is $\mathrm{o}$ for the photonic and $\mathrm{m}$ for the phononic subsystem in arm A and B, allows to identify the Bell-state, similarly to polarization entanglement in optical down-conversion experiments. Unlike the two mode entangled mechanical state in ~\cite{Riedinger2018}, this four-mode entangled optomechanical state allows us to realize a Bell measurement of the type suggested by Horne, Shimony and Zeilinger~\cite{Horne1989} and first demonstrated by Rarity and Tapster~\cite{Rarity1990} involving two-particle interference between four different modes. In order to access interferences between the mechanical modes, we convert the phonons into photons using a red detuned laser pulse (duration $\sim$40~ns, drive frequency $\nu_\mathrm{r}=\nu_\mathrm{o}-\nu_\mathrm{m}$). This realizes an optomechanical beamsplitter interaction which allows for a state transfer ($\op{H}{r} = -\hbar g_0 \sqrt{n_\mathrm{r}} \opd{a}{}\op{b}{}+\textrm{h.c.}$ , with the intracavity photon number $n_\mathrm{r}$). Note that this can also be described as a classical mapping process. The optical readout fields in the interferometer arms are again recombined on a beam splitter, after which the state of Stokes / anti-Stokes field is
\begin{eqnarray}
\ket{\Phi}{}&=&\frac{1}{2\sqrt{2}}[(1-e^{i(\phi_\mathrm{b}+\phi_\mathrm{r})})(\opd{a}{r1}\opd{a}{b1}-\opd{a}{r2}\opd{a}{b2})\nonumber\\
&+&i(1+e^{i(\phi_\mathrm{b}+\phi_\mathrm{r})})(\opd{a}{r1}\opd{a}{b2}+\opd{a}{r2}\opd{a}{b1})]\ket{0000}{}.
\label{eq:readout}
\end{eqnarray}

Here we express the detected fields in terms of their creation operators with labels $b$ ($r$) for photons scattered from the blue (red) drive and 1 (2) for the two detectors (cf.\ Fig.~\ref{fig:1}). Furthermore $\phi_\mathrm{r}$ is the phase difference that the red detuned pulse photons acquire in the two arms of the interferometer. Since experimentally the mechanical frequencies of the devices differ by a small offset $\Delta\nu_\mathrm{m}$ (see below), the state acquires an additional phase $\Omega=\Delta\nu_\mathrm{m}\Delta\tau$, where $\Delta\tau$ is the delay between the blue and red pulses. In all data below, however, we keep $\Delta\tau$ fixed such that we can treat it as constant and set $\Omega = 0$. Typically, Bell experiments are done by rotating the measurement basis in which each particle is detected. Equivalently, the state itself can be rotated, while keeping the measurement basis fixed. In our experiment we choose the latter option, as this is simpler to implement in our setup. We achieve this by applying a phase shift with an electro-optical modulator (EOM) in arm A of the interferometer, with which we can vary $\phi_{\mathrm{b}}$ and $\phi_{\mathrm{r}}$ independently (see SI). This allows us to select the relative angles between the photonic and phononic states.

In our experiment, the optical resonances are at a wavelength of $\lambda = 1550.4$~nm with a relative mismatch of $\Delta\nu_\mathrm{o}\approx150$~MHz. The mechanical modes have frequencies of $\nu_\mathrm{m}$ = 5.101~GHz and 5.099~GHz for device A and B, respectively. The bare optomechanical coupling rate $g_0/2\pi$ is 910~kHz for device A and 950~kHz for device B. While the optical mismatch is much smaller than the linewidth $\Delta \nu_\mathrm{o}\ll\kappa\sim 1$~GHz such that the devices are sufficiently identical, the mechanical mismatch requires optical compensation. This is realized using the EOM in arm A of the interferometer to ensure that the scattered photons from each arm interfere with a well defined phase on the second beamsplitter (see also SI).

At the base temperature of the dilution refrigerator of around 12~mK we obtain the phonon temperature of the mechanical modes by performing sideband asymmetry measurements~\cite{Riedinger2016}. The measured thermal occupations for both devices is $n_\mathrm{init}\le 0.09$. We determine the lifetimes of the phonons in our structures to be $\tau_\mathrm{A} = 3.3 \pm 0.5~\mu$s and $\tau_\mathrm{B} = 3.6 \pm 0.7~\mu$s using a pump-probe type experiment in which we excite the devices and vary the delay to the readout pulse. To re-initialize the devices in their groundstates prior to each measurement trial, we repeat the drive sequence every 50~$\mu$s, leaving more than 10 times their lifetime for thermalization with the environment. Furthermore, we set the delay between the blue and red detuned pulses to $\Delta\tau=200$~ns. The pulse energies for the Bell inequality experiment are chosen such that the excitation probability is 0.8\% (1\%), while the readout efficiency is 3\% (4.1\%) for device A (device B). These probabilities match the number of optomechanically generated photons for each device at the beamsplitter.

\begin{figure}[t!]
	\includegraphics[width=0.95\columnwidth]{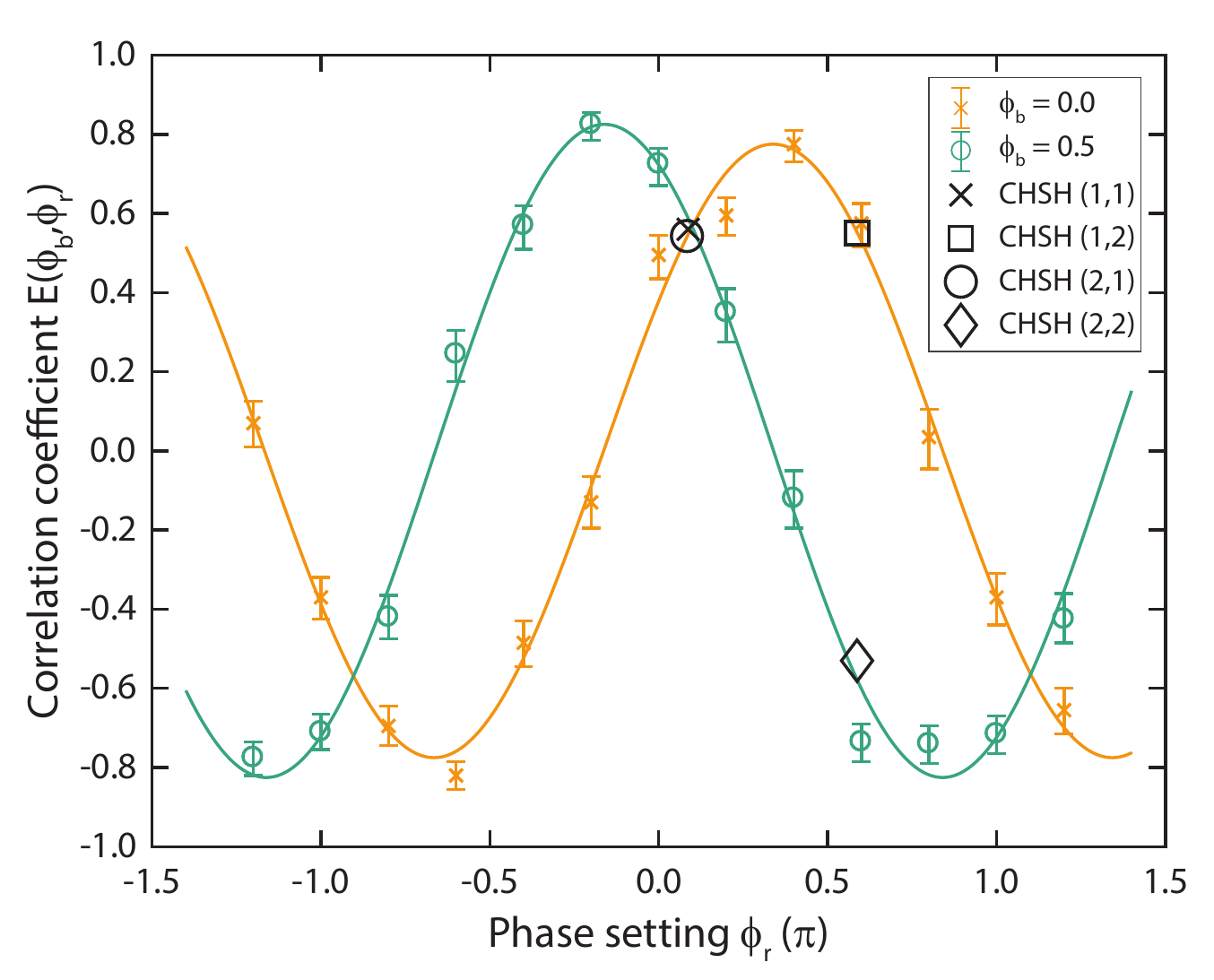}
	\caption{Correlation coefficients for various phase settings. We set the blue phase parameter $\phi_\mathrm{b}$ to 0 (orange) and $0.5\pi$ (green), while we scan the red pulse's phase setting $\phi_\mathrm{r}$ over more than $2\pi$. The optimal angles to test the CHSH inequality are shown with different symbols. The associated measured values can be found in Table~\ref{Tab:Bell}.
	\label{fig:2}}
\end{figure}

To characterize the performance of the devices, we first perform cross-correlation measurements of the photons scattered from blue and red drives on each individual optomechanical system. With the above mentioned settings, we obtain normalized cross-correlation values of $g^{(2)}_{\mathrm{br,A}}$ = 9.3 $\pm$ 0.5 and $g^{(2)}_{\mathrm{br,B}}$ = 11.2 $\pm$ 0.6~\cite{Riedinger2016}. We can use this to estimate the expected interferometric visibility for the experiments below as $V_\mathrm{xpcd}=\frac{g^{(2)}_{\mathrm{br}}-1}{g^{(2)}_{\mathrm{br}}+1}$~\cite{Riedmatten2006}. As there is a small mismatch in the observed cross-correlations of the two devices, we use the smaller value of device A, which results in an expected visibility of around $V_\mathrm{xpcd}=81\%$.

In order to experimentally test a Bell inequality, we then drive the two devices simultaneously in a Mach-Zehnder interferometer (see Fig.~\ref{fig:1} and SI). We define the correlation coefficients
\begin{equation}\label{eq:CorrCoeff}
E(\phi_{\mathrm{b}},\phi_{\mathrm{r}})=\frac{n_{11}+n_{22}-n_{12}-n_{21}}{n_{11}+n_{22}+n_{12}+n_{21}}.
\end{equation}
Here $n_{ij}$ represents the number of detected coincidences scattered from blue ($i$) and red ($j$) pulses on the two detectors ($i,j=1,2$), such that e.g.\ $n_{21}$ is the number of trials where the blue drive resulted in an event on detector 2, whereas the consecutive red drive on detector 1. The visibility $V$ is given as the maximum correlation coefficient $V=\rvert E(\phi_{\mathrm{b}},\phi_{\mathrm{r}})\rvert_{\mathrm{max}}$. We measure the correlation coefficients for various phase settings for the blue ($\phi_{\mathrm{b}}$) and red ($\phi_{\mathrm{r}}$) pulses, as shown in Figure~\ref{fig:2}. Strong correlations in the detection events by photons scattered from blue and red pump pulses can be seen, of which the latter are a coherent mapping of the mechanical state of the resonator. This sweep demonstrates that we are able to independently shift the phases for the Stokes and anti-Stokes states. The visibility $V=80.0\pm2.5\%$ we obtain from fitting the data matches the prediction from the individual cross-correlation measurements very well. The interference furthermore shows the expected periodicity of $2\pi$.

\begin{table}
	\centering
	\renewcommand*{\arraystretch}{1.4}
	\begin{tabular}{C{0.2\columnwidth} C{0.1\columnwidth} C{0.1\columnwidth} C{0.4\columnwidth}}
		\hline
			Settings $i,j$ & $\phi^i_\textrm{b} [\pi]$ & $\phi^j_\textrm{r} [\pi]$ & $\hspace{8pt}E(\phi_\textrm{b},\phi_\textrm{r})$ \\ \hline
			(1,1) & 0.0 & 0.087 & $\hspace{6pt}0.561_{-0.020}^{+0.019}$ \\ 
			(1,2) & 0.0 & 0.587 & $\hspace{6pt}0.550_{-0.022}^{+0.020}$ \\ 
			(2,1) & 0.5 & 0.087 & $\hspace{6pt}0.542_{-0.021}^{+0.018}$ \\	
			(2,2) & 0.5 & 0.587 & $-0.523_{-0.021}^{+0.021}$ \\ 

			\hline
	\end{tabular}
	\caption{Correlation coefficients for the optimal CHSH angles. The violation of the inequality can be computed according to Eq.~\eqref{eq:CHSH} and results in a $S$ value of $S=2.174_{-0.042}^{+0.041}$, corresponding to a violation of the classical bound by more than 4 standard deviations.}
	\label{Tab:Bell}
\end{table}

To test possible local hidden-variable descriptions of our correlation measurements we use the CHSH-inequality~\cite{Clauser1969}, a Bell-type inequality. Using the correlation coefficients $E(\phi_{\mathrm{b}},\phi_{\mathrm{r}})$, it is defined as
\begin{equation}
S=\rvert E(\phi^{1}_{\mathrm{b}},\phi^{1}_{\mathrm{r}})+E(\phi^{1}_{\mathrm{b}},\phi^{2}_{\mathrm{r}})+E(\phi^{2}_{\mathrm{b}},\phi^{1}_{\mathrm{r}})-E(\phi^{2}_{\mathrm{b}},\phi^{2}_{\mathrm{r}})\rvert\leq 2.
\label{eq:CHSH}
\end{equation}
A violation of this bound allows us to exclude a potential local realistic theory from describing the optomechanical state that we generate in our setup. The maximal violation $S_{QM}=2\sqrt{2}\cdot V$ is expected for settings $\phi^i_{\mathrm{b}}=[0,\pi/2]$ and $\phi^j_{\mathrm{r}}=[-\pi/4+\phi_\textrm{c},\pi/4+\phi_\textrm{c}]$, with $i,j=1,2$~\cite{Cirelson1980}. Here $\phi_\textrm{c}=0.337\pi$ is an arbitrary, fixed phase offset that is inherent to the setup. Our experimentally achieved visibility exceeds the minimal requirement for a violation of the classical bound $V \geq 1/\sqrt{2}\approx 70.7\%$. We proceed to directly measure the correlation coefficients in the four settings, as indicated in Figure~\ref{fig:2}, and obtain $S=2.174_{-0.042}^{+0.041}$ (cf.\ Table~\ref{Tab:Bell}). This corresponds to a violation of the CHSH inequality by more than 4 standard deviations, clearly confirming the non-classical character of our state. From the observed visibility of $V=80.0\%$, we would expect a slightly stronger violation with $S\approx 2.26$. The reduction in our experimentally obtained value for $S$ can be attributed to imperfect filtering of drive photons in front of one of the SNPSDs, which gives rise to varying amounts of leak photons at different phase settings (see discussion in SI).\\

\begin{figure}[t!]
	\includegraphics[width=0.95\columnwidth]{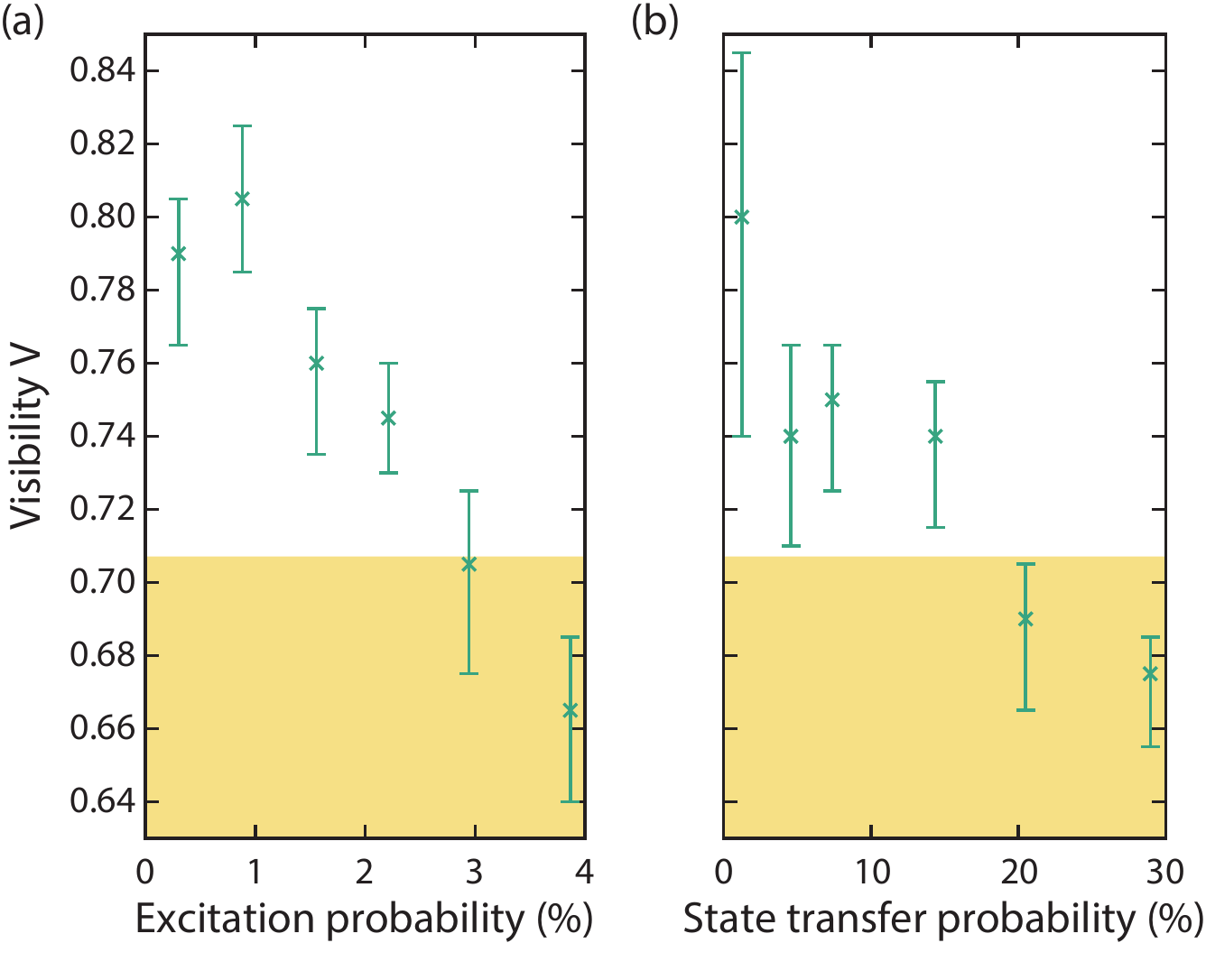}
	\caption{Visibility as a function of generation rate and state transfer probability. We sweep the power of the blue pulse while keeping the red state transfer probability fixed, inducing absorption of the optical field in the silicon structure, and see that for excitation probabilities up to around 3\% the measured visibility exceeds the threshold to violate the CHSH inequality (left). When increasing only the red pump power (right) a similar behavior can be observed, allowing us to increase the state transfer probability beyond 14\%, while still being able to overcome the classical bound (orange shaded region). The visibilities $V=\rvert E(\phi_{\mathrm{b}},\phi_{\mathrm{r}})\rvert_{\mathrm{max}}$ are measured in a single phase setting at the optimal angles $\phi_{\mathrm{b}}=0$ and $\phi_{\mathrm{r}}=0.337\pi$.}
	\label{fig:3}
\end{figure}

For quantum network applications it is also important to analyze the quality of the detected optomechanical entanglement with regard to the detection rate. In our measurements we can achieve this by changing the energies of the drive beams to alter the optomechanical interaction strengths. An increase in the blue pulse energy is accompanied by two mechanisms that decrease the state fidelity. Firstly, the probability for higher order scattering events $\mathcal{O}(p_\mathrm{A,B}^2)$ is increased. Secondly, higher pulse energies also result in more absorption, degrading the state through thermal excitations. As observed in previous experiments~\cite{Riedinger2016,Riedinger2018}, optical  pumping of the devices creates a thermal population of the mechanical modes with timescales on the order of several hundreds of nanoseconds (see also SI). While we keep the delay to the readout pulse short ($\Delta\tau = 200$~ns), we cannot fully avoid these spurious heating effects. Hence the decrease in visibility with increased pulse energy, as seen in Figure~\ref{fig:3}(a), can be attributed mostly to this direct absorption heating. To further test the heating dynamics of our state, we also sweep the red pulse energies while keeping the excitation energy fixed at the value used in the main experiment ($p_\mathrm{b}=$ 0.8\% and 1\%). As expected, the increased readout pulse energies lead to substantial heating of the devices~\cite{Hong2017}. However, even for relatively large optical powers corresponding to $\sim$14\% read out efficiency, the correlation coefficient is above the threshold for violating a Bell inequality under the fair sampling assumption, see Figure~\ref{fig:3}(b).

Our system is fully engineered and hence we have complete control over the resonance frequencies and possibilities to integrate with other systems. While in our current structures we intentionally cap the mechanical quality factors to keep the measurement time short~\cite{Patel2017}, recent experiments with very similar devices have observed lifetimes larger than 1~s~\cite{MacCabe2018}. Long lived non-classical states of large masses are interesting for fundamental studies of quantum mechanics. Combined with the fact that we can efficiently couple these states to photons in the telecom band could enable interesting experiments with Bell tests at remote locations. Employing fast optical switches that route one of the photons to a second set of detectors would furthermore allow us to close the locality loophole in the future. Our probabilistic scheme could, in principle, also be adapted to perform a 'loophole-free' Bell test~\cite{Giustina2015}, if in addition the detection loophole would be closed through a more efficient read-out.

In summary, we have demonstrated the violation of a Bell-type inequality using massive (around $10^{10}$ atoms), macroscopic optomechanical devices, thereby verifying the non-classicality of their state without the need for a quantum description of our experiment. The experimental scheme demonstrated here may also be employed in other, even more massive optomechanical systems. One outstanding challenge is to generate states of genuine macroscopic distinction, for example a macroscopic separation in the center of mass, to investigate fundamental decoherence mechanisms~\cite{Bassi2013} or even the interplay between quantum physics and gravity~\cite{Bose2017,Marletto2017}. We also show that the created entangled states are relatively robust to absorption heating, which could lead to a realistic implementation of entanglement generation for a future quantum network using optomechanical devices. Violation of a CHSH inequality can also be used to verify long-distance quantum communication with device-independent security using mechanical systems.

\begin{acknowledgments}
We would like to thank Vikas Anant, Nicolas Sangouard, and Joshua Slater for valuable discussions and support. We also acknowledge assistance from the Kavli Nanolab Delft, in particular from Marc Zuiddam and Charles de Boer. This project was supported by the Foundation for Fundamental Research on Matter (FOM) Projectruimte grants (15PR3210, 16PR1054), the European Research Council (ERC StG Strong-Q, ERC CoG QLev4G), the European Commission under the Marie Curie Horizon 2020 initial training programme OMT (grant 722923), the Vienna Science and Technology Fund WWTF (ICT12-049), the Austrian Science Fund (FWF) under projects F40 (SFB FOQUS) and P28172, and by the Netherlands Organisation for Scientific Research (NWO/OCW), as part of the Frontiers of Nanoscience program, as well as through a Vidi grant (680-47-541/994). R.R. is supported by the FWF under project W1210 (CoQuS) and is a recipient of a DOC fellowship of the Austrian Academy of Sciences at the University of Vienna.
\end{acknowledgments}

\setcounter{figure}{0}
\renewcommand{\thefigure}{S\arabic{figure}}
\setcounter{equation}{0}
\renewcommand{\theequation}{S\arabic{equation}}

\clearpage

\section{Supplementary Information}

\subsection{Experimental setup}

\begin{figure}[t]
		\includegraphics[width=1.0\columnwidth]{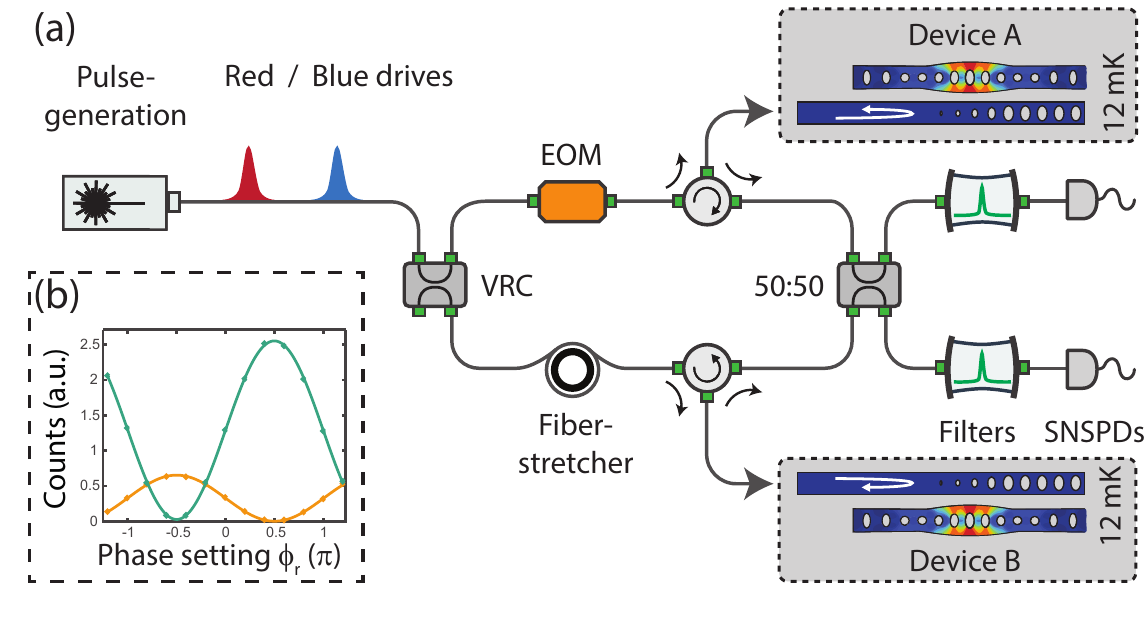}
		\caption{(a) Sketch of the setup. We first generate blue and red pulses that drive the optomechanical interactions. The devices are each placed in one arm of a fiber interferometer realized between a variable ratio coupler (VRC) and a 50:50 beamsplitter. We stabilize the overall phase with a fiber stretcher such that we can select particular phase settings with our EOM in arm A. The optomechanically generated photons are filtered from the drive pulses before finally being detected on single-photon superconducting nanowire detectors (SNSPDs). (b) Interferometer visibility. We detune the filter locking point from the device resonance such that we are only sensitive to pump photons that leak through the filters. Here we plot the rate of pump photons detected with detector~1 (orange) and detector~2 (green) as a function of phase difference of the interferometer arms. We observe the expected $2\pi$ periodicity with an interference visibility of 98.4\%. The difference in amplitude stems from different degrees of suppression of pump photons through the two filter lines.}
		\label{fig:S1}
\end{figure}

A sketch of the fiber-based setup used in the main text is shown in Figure~\ref{fig:S1}. The pulse generation consists of two tunable diode lasers (Santec TSL550 and Toptica CTL1550), which are stabilized at the sidebands of device~B using a wavelength meter. We suppress high frequency noise on both laser through optical filtering (linewidth $\sim$50~MHz), before we generate the drive pulses using acousto-optic modulators (pulse length $\sim$40~ns). The interferometer is formed by a variable ratio coupler and a calibrated 50:50 coupler (deviation below 3\%). The interferometer has a free spectral range of 1.2~GHz and is phase-stabilized with a home built fiber stretcher. The EOM is used to select a desired phase on a fast timescale and simultaneously to compensate the frequency mismatch of the mechanical devices of $\Delta\nu_\mathrm{m}=2.3$~MHz. This mismatch is small enough to be compensated by a linear phase sweep during the pulses without the need of a serrodyne drive.

In order to achieve high efficiencies in our detection paths we use a home-built freespace filtering setup. Each filter line consists of two linear cavities which are actively stabilized to the resonance of the devices. The total detection efficiency for optomechanically scattered photons from device~A is 3.4\% for detector~1 and 2.9\% for detector~2. The efficiency for device~B is 2.9\% for detector~1 and 2.3\% for detector~2. The total loss budget consists of various contributions:\ photons that are created in one of the optomechanical cavities are transfered to an on-chip silicon waveguide with efficiencies of 65\% and 55\% for devices A and B, respectively. The transmission from the waveguide to the output of the circulator (59\% and 55\%) is dominated by waveguide to fiber coupling losses. The rest of the losses are due to a finite transmission through filters, optical components needed for feeding continuous locking light and finite detection efficiency of SNSPDs.

In order to evaluate the quality of our interferometer we record the first order interference of our lasers. For this, we detune the filters by $\sim$2~GHz from the optomechanically scattered photons, such that we are only sensitive to leaked pump photons. We then lock the interferometer with the fiber stretcher and sweep the phase using the EOM as we do for the visibility sweep in Figure~\ref{fig:1}. The visibility we obtain of 98.4\% matches well with the independently measured short term fluctuations of the interferometer lock of around $\sim$$\pi/25$~\cite{Minar2008}. The main cause of these fluctuations is noise that is picked up by the fibers inside the dilution refrigerator stemming from the pulse tube cryo cooler.

\subsection{Pump filtering and false coincidences}

To estimate the effect of erroneous coincidence clicks that do not stem from the optomechanical state, we perform calibration measurements to estimate the role of leaked drive photons. To do this, we slightly detune the filters away from the frequency of the optomechanicaly scattered photons, such that they are reflected from the filters and do not reach the SNSPDs. We find that during the main experiments around 17\% of red counts detected on detector~2 are in fact imperfectly filtered drive photons. For detector~1 this number is around 7\%. To understand this asymmetry, we note that the cavities in front of detector~1 both have linewidths of $\sim$35~MHz, while the ones in front of detector~2 have a slightly larger linewidth of $\sim$45~MHz. We estimate that perfect filtering would enable us to obtain roughly 12\% higher cross correlation values for the individual devices. Similarly, we measure that less than 2\% of the detected photons during the blue pulses are leak photons.

The asymmetry in pump suppression has additional consequences for the experiments in which we drive the devices simultaneously. The total rate of leak photons varies with the selected phase setting $\phi_\textrm{r}$. The angular dependency is proportional to the sum of the two curves in inset (b) of Figure~\ref{fig:S1}. These unwanted photons result in additional coincidences in the second order interference of the main experiment, hence they distort the visibility sweeps of Figure~\ref{fig:2}. The purely sinusoidal fits are not capturing this accurately and therefore mostly serve as a guide to the eye. The values of the correlation coefficients $E(\phi_\textrm{b},\phi_\textrm{r})$ in Table~\ref{Tab:Bell} are affected in the same way. We measure at angles $\phi_\textrm{r}=0.087\pi$ and $0.587\pi$, of which the latter suffers more from the imperfectly filtered drive pulses. This is the main reason of why we observe a reduction in the Bell parameter $S$ compared to the expected results from the visibility $V$ in figure~\ref{fig:2} alone. Dark counts on the other hand are low enough (around 15~Hz) to only contribute by less than 1\% to the detection events.

\begin{figure}[t]
		\includegraphics[width=1.0\columnwidth]{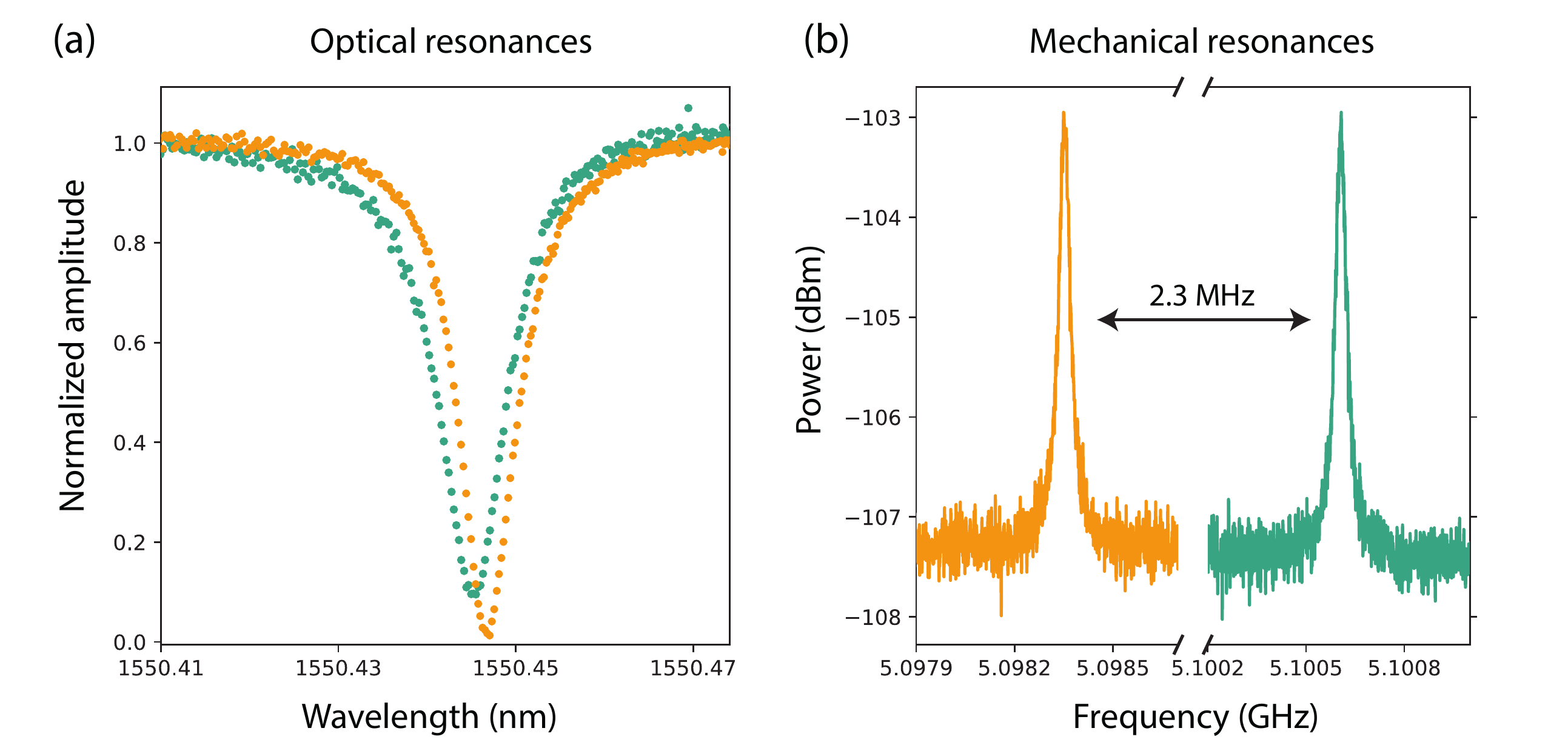}
		\caption{Spectroscopy of optomechanical devices A (green) and B (orange). Left:\ Optical cavity reflection spectrum of the cavities. Right:\ Mechanical resonances of the two devices.}
		\label{fig:S2}
\end{figure}

\subsection{Device fabrication and characterization}

The optomechanical devices are fabricated from silicon-on-insulator wafers with a device film thickness of 250~nm as described in~\cite{Hong2017}. In order to reliably find identical devices on distinct chips, we optimized the electron-beam doses throughout the lithography step, which allowed us to obtain a distribution of only 1~nm of the optical resonances. To further reduce the variability between different chips, we first fabricate a single large chip, fully process it, and then cleave it into smaller pieces as a last step. This optimized procedure results in two chips with excellent overlap of the optical resonances. Unlike in previous experiments~\cite{Riedinger2018}, our fabrication method allowed us to perform the experiments without the need for serrodyne shifting of the photon frequencies in one of the arms of the interferometer but rather only apply a small linear ramp signal to the red and blue pulses using the EOM in arm A.

We characterize the optical resonances by sweeping a continuous-wave part of our laser and recording the reflected intensity resonances (cf.\ Figure~\ref{fig:S2}(a)). As we use reflectors at the end of our waveguides (see Figure~\ref{fig:1}), we effectively couple in a single sided way to our devices and therefore expect to see resonances as dips in the reflected light. The measurements of the mechanical resonances are performed by locking the laser to the blue sideband of the optical resonances, amplifying the reflected light in a fiber amplifier and then detecting the optomechanically generated sideband on a fast photodiode (Figure~\ref{fig:S2}(b)).

\begin{figure}[t]
		\includegraphics[width=1.0\columnwidth]{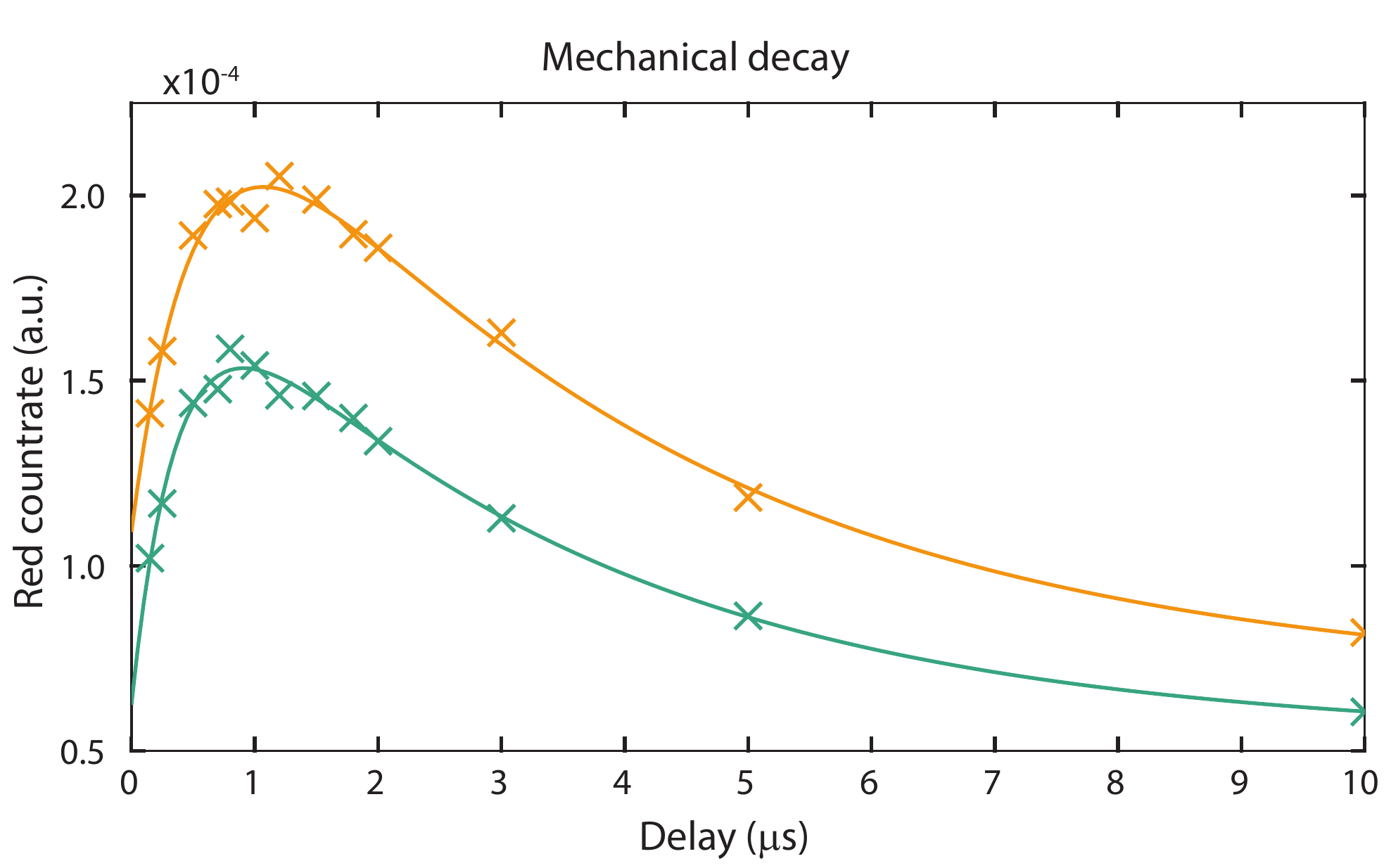}
		\caption{Absorption heating and mechanical decay. Heating dynamics of device A (green) and device B (orange) from a pump-probe experiment are plotted. We excite the devices with a blue pulse and read out the mechanical state after a variable delay $\mathrm{\Delta}\tau$. Thermal phonons due to absorption of the blue drive occupy the mode on a timescale of several hundreds of nanoseconds, after which the excitations decay to the cold environment with the mechanical lifetimes.
		\label{fig:S3}}
\end{figure}

To evaluate the absorption heating dynamics, we perform a pump-probe experiment with the individual devices. Here we excite the devices with a blue and probe with a red drive pulse after a variable delay $\Delta\tau$, see Figure~\ref{fig:S3}. The pulse energies are chosen similarly to the ones in the main text. As already observed in previous work~\cite{Riedinger2016,Hong2017,Riedinger2018}, the devices see absorption caused heating with a rise time of several hundreds of nanoseconds. The highest temperature is reached in both cases at a delay of around 1~$\mu$s, after which the devices decay with their intrinsic quality factors. We fit the excitation dynamics $d_i(\Delta\tau)$ of the two devices ($i=$ A, B) with a phenomenologically motivated double exponential model of the form $d_i(\Delta\tau) = a_i e^{-\Delta\tau/\tau_i} - b_i e^{-\Delta\tau/\eta_i} + n_{\mathrm{init},i}$~\cite{Riedinger2018}. We can extract the energy lifetimes of our devices as $\tau_\mathrm{A} = 3.3 \pm 0.5~\mu$s and $\tau_\mathrm{B} = 3.6 \pm 0.7~\mu$s. Note that the final decay level $n_{\mathrm{init},i}$ does not represent the true base temperature, as we still observe counts from the intra-pulse heating. Nevertheless, we can bound the occupancy $n_{\mathrm{init},i}$ from above using the asymmetry in the click rates of blue and red sideband scattered photons $C_{\mathrm{b},i}$ and $C_{\mathrm{r},i}$ as $n_{\mathrm{init},i} = C_{\mathrm{r},i} / (C_{\mathrm{b},i} - C_{\mathrm{r},i})$~\cite{Riedinger2016}. We perform the measurements for blue and red pulses individually with a duty cycle of 50~$\mu$s each. For device~A, we determine the initial occupation to be $\sim$0.07 phonons if measured on detector~1 and $\sim$0.09 phonons if measured on detector~2. Device~B has similar apparent occupations of $\sim$0.06 phonons on detector~1 and $\sim$0.09 phonons if measured on detector~2. The difference in the measurements on the two detectors reflects the different amount of leak suppression as discussed above. However, both measurements also contain the intrapulse heating detection events, meaning we expect the real occupancy in the dark to be below these extracted numbers.

\subsection{Statistical analysis}

The statistical analysis for the CHSH inequality is done using the same techniques as Ref.~\cite{Riedinger2018}. We apply binominal statistics on the number of coincidence events $n_{ij}$ in Equation~\eqref{eq:CorrCoeff} and generate discrete probability distributions. We then treat the correlation coefficients $E$ as non-trivial functions of two random variables ($n_\mathrm{same} = n_{11}+n_{22}$ and $n_\mathrm{diff} = n_{12}+n_{21}$) and numerically find their probability distributions via the cummulative density function method. We calculate the probability distribution for the Bell parameter $S$ as a convolution of the probability distributions of the four correlation coefficients. Finally we calculate the expectation value and error bounds ($\pm$34\% confidence interval) by numerical integration of the resulting probability density function.\\

The data for the main experiment was acquired in with approx.\ $500-620$ million trials per CHSH setting. Together with the duty time of 50~$\mu$s, this amounts to pure measurement times of $7.0-8.6$ hours per setting, or $\sim$31 hours in total. This time is excluding additional overhead that is needed to re-lock the filters or manage the acquired data. The total actual measurement time is about a factor of two larger. The data was taken in intervals of 20 minutes and the CHSH settings were cycled after each interval. During the experiment, a total of 6423 photon pairs were detected, which amounts to roughly 210 successful trials per pure measurement hour.

\begin{table}
	\centering
	\begin{tabular}{ccccccc}

		\hline
			CHSH setting & Trials & Heralding clicks & $n_{00}$ & $n_{01}$ & $n_{10}$ & $n_{11}$ \\ \hline

			(1,1) &  597302527 & 645858 & 708 & 194 & 175 & 611 \\
			(1,2) &  500363903 & 546488 & 606 & 162 & 164 & 521 \\
			(2,1) &  622224596 & 680260 & 752 & 212 & 185 & 589 \\
			(2,2) &  540137661 & 592728 & 170 & 586 & 590 & 198 \\

			\hline
	\end{tabular}
	\caption{Recorded coincidence clicks of the main experiment. Individual trials are performed every 50~$\mu$s. Heralding clicks are the detected photons scattered by the blue pulse. Coincidences $n_{i,j}$ are clicks that were registered after getting a heralding event in the same trial, with $i,j$ indicating the detector for the blue and red scattered photons.}
	\label{Tab:Clicks}
\end{table}

\end{document}